\documentclass[english, ]{elsarticle}
\usepackage[T1]{fontenc}
\usepackage[latin9]{inputenc}
\usepackage{graphicx}
\usepackage{setspace}
\usepackage{amssymb}
\usepackage{esint}
\usepackage{babel}

\begin{document}

\title{Infrared absorption study of charge ordered $La{}_{0.5}Ca{}_{0.5-x}Sr_{x}MnO_{3}$
$(0.1\leq x\leq0.5)$ manganites}

\author{Indu Dhiman}

\address{Solid State Physics Division, Bhabha Atomic Research Centre, Mumbai
- 400~085, India}

\author{A. Das}

\ead{adas@barc.gov.in}

\address{Solid State Physics Division, Bhabha Atomic Research Centre, Mumbai
- 400~085, India}

\author{K. R. Priolkar}

\address{Department of Physics, Goa University, Taleigao Plateau, Goa - 403~206,
India}

\author{P. S. R. Murthy}

\address{Department of Physics, Goa University, Taleigao Plateau, Goa - 403~206,
India}
\begin{abstract}
Infrared absorption study has been carried out on a series of half
doped manganites, $La_{0.5}Ca_{0.5-x}Sr_{x}MnO_{3}$ $(0.1\leq x\leq0.5)$,
with varying magnetic ground state. The charge ordering transition
observed in samples with {\normalsize $x\leq0.3$ is accompanied by
a mode at $\sim525cm^{-1}$ in addition to the stretching mode at
$615cm^{-1}$ and bending mode at $400cm^{-1}$. Phonon hardening
is found to occur below the CE - type antiferromagnetic ordering temperature.
The value of the insulating gap decreases on doping with Sr from $727cm^{-1}$
to $615cm^{-1}.$}{\normalsize \par}\end{abstract}
\maketitle

\section{Introduction}

The correlation of magnetic and transport behavior in half doped
$R{}_{0.5}A{}_{0.5}MnO{}_{3}$ (R: a trivalent earth ion, A: a
divalent earth ion) manganites have lead to considerable research
interest in these systems \cite{C N R Rao,Dagotto,J. B.
Goodenough,Y. Tokura}. Such correlated behavior in manganites has
been explained by the interplay of double exchange and super
exchange interactions coupled with strong electron - phonon
interactions arising due to breathing and Jahn-Teller modes
\cite{A. J. Millis}. The Jahn-Teller distortion and double
exchange interactions are competitive in nature. While the former
tends to localize the charge in distortions of oxygen octahedra
around $Mn{}^{3+}$ ions, latter favors charge delocalization
\cite{D. Louca,C. P. Adams}. As a consequence, the distortions of
$Mn{}^{3+}O{}_{6}$ octahedra strongly influences the evolution of
magnetic and structural phases. The infrared phonon spectra are
sensitive to these local lattice distortions and its study sheds
light on the important role played by $MnO_{6}$ vibrations, across
metal-insulator transition \cite{K. H. Kim}.

The $LaMnO{}_{3}$ compound is an antiferromagnetic insulator
having an orthorhombic structure in \textit{Pnma} space group. The
dominant Jahn-Teller effect arising from the presence of
$Mn{}^{3+}$ ions causes the structure to deviate from cubic to
orthorhombic \cite{A. J. Millis-1}. Infrared absorption study of
this compound show 10 out of 25 lines predicted under orthorhombic
symmetry \cite{M. N. Iliev}. In the mid infrared range this
compound shows negligible absorption at all temperatures.
Comparative analysis of measured spectra in association with
lattice dynamical calculations allow reasonable assignment for the
phonon modes observed in $LaMnO{}_{3}$ \cite{M. Rini,E. G. Rini}.
Doping rare-earth with an alkaline earth metals like Ca, Sr, or
Ba, in $R{}_{1-x}A{}_{x}MnO{}_{3}$ (A: alkaline earth metals)
results in conversion of an appropriate number of $Mn{}^{3+}$ to
$Mn{}^{4+}$ ions. This gives rise to ferromagnetic
$Mn{}^{3+}-O-Mn{}^{4+}$ double exchange interactions whereby
$e_{g}$ electrons from $Mn{}^{3+}$ is transferred to $Mn{}^{4+}$
with a parallel spin configuration. The Ca doped
$La{}_{1-x}Ca{}_{x}MnO{}_{3}$ manganites display a complex phase
diagram as a function of doping (x) and temperature. Of particular
interest are the half doped compounds which display coexisting
ferromagnetic, antiferromagnetic, charge and orbital ordering as a
function of temperature. The charge and orbital ordering in these
systems is accompanied by a change in the sound velocity
indicating a strong electron - acoustic phonon coupling \cite{J.
D. Lee}. The acoustic phonon frequency in this case is shown to be
dependent upon the magnetic susceptibility in the
antiferromagnetic state. This indicates a coupling between the
phonons and magnetization which could be explored by infrared
measurements. The Jahn - Teller mode, which plays an important
role in the magnetism of hole doped manganites, is influenced by
the optical phonon frequency \cite{J. D. Lee-1}. Indeed, in the
study of infrared spectra in $La_{1-x}Ca_{x}MnO_{3}$, it is found
that the stretching modes progressively shifts from 596 to
588$cm^{-1}$ on increasing Ca from $LaMnO_{3}$ to $CaMnO_{3}$
\cite{L. Kebin}. In another study, pressure dependent infrared
studies on $La_{0.8}Ca_{0.2}MnO_{3}$ compound show the closing of
optical gap with increasing pressure leading to metallization
\cite{A. Sacchetti}. Kim et al. reported the external, bending,
and stretching modes frequencies of $MnO_{6}$ octahedra in
$La_{0.7}Ca_{0.3}MnO_{3}$ compound \cite{K. H. Kim}. Infrared
absorption studies have been reported for charge and orbitally
ordered $La{}_{0.5}Ca{}_{0.5}MnO{}_{3}$ compound \cite{P.
Calvani}. This compound exhibits ferromagnetic transition at
$T_{C}\approx225K$ and an antiferromagnetic transition at
$T_{N}\approx170K$, in addition to the transition from
incommensurate to commensurate charge ordering in the
antiferromagnetic region \cite{P. G. Radaelli,C. H. Chen}. The
previously reported infrared absorption study in
$La{}_{0.5}Ca{}_{0.5}MnO{}_{3}$ compound show the formation of an
optical gap on lowering of temperature, leading to localization of
the carriers \cite{P. Calvani}. As the screening action of the
mobile holes weakens, phonon peaks become increasingly evident at
lower temperature. However, the previous studies are inconclusive
on the variation of infrared active modes in half doped compounds
with varying magnetic ground states.

In the present study, we have made an attempt to understand the
correlated variations in the infrared active modes of vibration of
$MnO_{6}$ octahedra in association with polaronic absorption
background, in charge and orbitally ordered systems. Towards this,
temperature dependent infrared absorption spectroscopy study in
$La_{0.5}Ca_{0.5-x}Sr_{x}MnO_{3}$ $(0.1\leq x\leq0.5)$ manganites
has been carried out. Previously reported neutron diffraction
studies on these compounds show that with Sr doping, the CE
\textendash{} type antiferromagnetic structure is stable for x <
0.3. At x = 0.3, CE \textendash{} type ordering coexists with an A
\textendash{} type antiferromagnetic phase. At x = 0.4, A
\textendash{} type antiferromagnetic order replaces the CE
\textendash{} type state, in addition with the evidence of long
range ferromagnetic ordering in the temperature range of 180
\textendash{} 250K. All the compounds with $x\leq0.4$ exhibit an
insulating behavior. In x = 0.5 composition, long range
ferromagnetic metallic phase is observed at temperatures below
310K \cite{I. Dhiman,I. Dhiman-1}. Pressure dependent neutron
diffraction studies in these compounds show the equivalence of
increasing $<r_{A}>$ and increasing external pressure in these
compounds \cite{I. Dhiman-2}.

\section{Experiment}

The polycrystalline samples used for these measurements are the
same on which neutron diffraction studies have been carried out
and reported earlier \cite{I. Dhiman}. Infrared measurements were
performed using Shimadzu FTIR-8900 spectrophotometer at various
temperatures below 300K in transmission mode in the range of 400
to 4000 $cm^{-1}$. Samples were mixed with KBr in the ratio of
1:100 by weight and pressed into pellets for infrared
measurements. Temperature variation was achieved using an Oxford
optical cryostat with KRS5 windows.

\section{Results and Discussion}

The structural and magnetic phase diagram of
$La_{0.5}Ca_{0.5-x}Sr_{x}MnO_{3}$ $(0.1\leq x\leq0.5)$ series
obtained from unpolarized and polarized neutron diffraction study
has been reported previously \cite{I. Dhiman,I. Dhiman-1}. The
Sr-doped compounds in the $La_{0.5}Ca_{0.5-x}Sr_{x}MnO_{3}$ series
with $x\leq0.3$ are isostructural, possessing an orthorhombic
distorted perovskite structure in \textit{Pnma} space group at
300K. On lowering of temperature, these compounds exhibit
transition to charge and orbital ordering at
$T_{CO}\approx225-250K$. The transition to a charge ordered state
below $T_{CO}$ is accompanied by a change of symmetry from
orthorhombic structure in \textit{Pnma }space group to monoclininc
structure in \textit{$P2_{1}/m$} space group. The x = 0.4 sample
crystallizes with two orthorhombic phases in the space group
\textit{Pnma} and \textit{Fmmm}, whereas the end composition at x
= 0.5 has a tetragonal structure in \textit{I4/mcm} space group.
Magnetically, with progressive increase in Sr doping, the CE
\textendash{} type antiferromagnetic phase is suppressed and
ferromagnetic phase is favored. The CE \textendash{} type
antiferromagnetic phase is observed in x = 0.1 and 0.3 samples
with antiferromagnetic transition temperatures as 175 and 150K,
respectively. In x = 0.4, the CE \textendash{} type
antiferromagnetic structure is fully suppressed and A
\textendash{} type antiferromagnetic phase is observed with the
transition temperature $T_{N}$ = 200K. In addition, evidence of
long range ferromagnetic ordering in the temperature range of 180
to 250K is observed only in x = 0.4 sample. Further increase in Sr
doping to x = 0.5, the long range ordered ferromagnetic phase is
established at all temperatures below 310K. In the present study
we have selected few samples, such that each displays a different
magnetic ground state.

\begin{figure}
\includegraphics{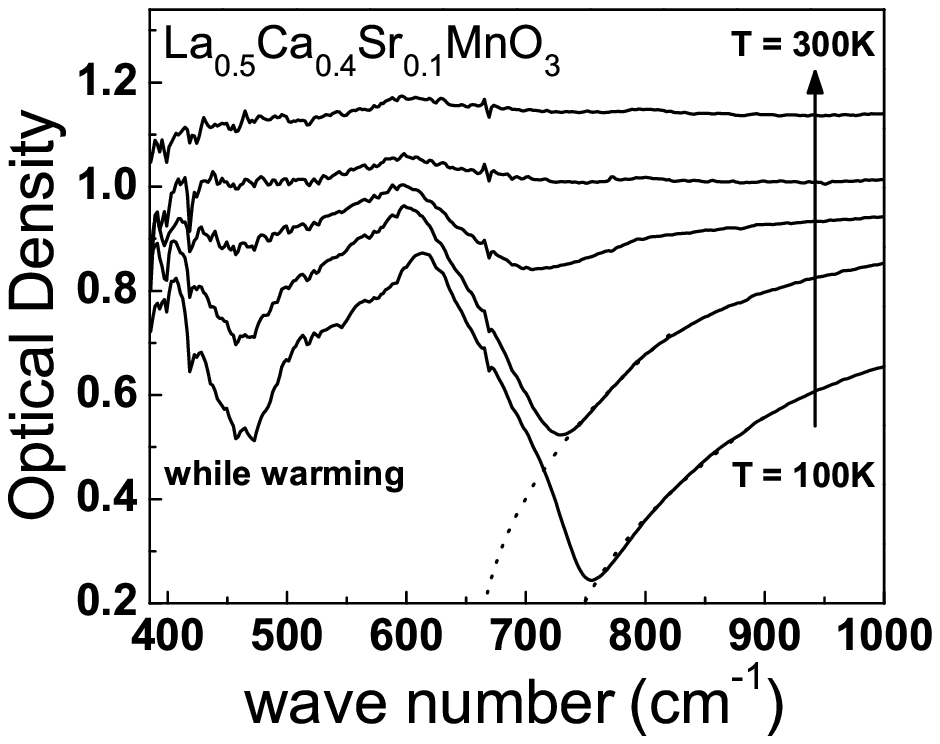}

\caption{\label{Optical-density Sr01} Optical density $O_{d}(\omega)$ for
$La{}_{0.5}Ca{}_{0.4}Sr{}_{0.1}MnO_{3}$ (x = 0.1) sample at various
temperatures between 100 to 300K. The dotted lines represent the extrapolation
based of equation \ref{eq:optical conductivity}. Optical density
patterns have been artificially shifted to enhance clarity.}

\end{figure}

The optical density ($O_{d}(\omega)$) in the infrared range
between 400 and 1500$cm{}^{-1}$ of
$La{}_{0.5}Ca{}_{0.4}Sr{}_{0.1}MnO_{3}$ (x = 0.1) sample at
temperatures between 100 and 300K is shown in figure
\ref{Optical-density Sr01}. The optical density $O_{d}(\omega)$,
which is proportional to the optical conductivity
$\sigma(\omega)$, is defined as \cite{A. Paolone},

\[
O_{d}(\omega)=ln[I_{KBr}(\omega)/I_{s}(\omega)]\varpropto\sigma(\omega)\]

\noindent where, $I_{KBr}(\omega)$ and $I_{s}(\omega)$ are the
infrared intensities transmitted by a pure KBr pellets with and
without the sample, at the same temperature, respectively. At 300K
in x = 0.1 sample, two infrared absorption peaks at $\sim400$ and
$600$ $cm^{-1}$ are observed. In \textit{Pnma} space group there
are 25 infrared active modes of vibrations,
$9B_{1u}+7B_{2u}+9B_{3u}$. The infrared active phonon modes of
symmetry $B_{3u}$, $B_{2u}$, and $B_{1u}$ indicate toward the
oscillations of dipole moment along the x, y, and z directions,
respectively. Additionally, there are $8A_{u}$ infrared silent
modes and 3 acoustic modes as $B_{1u}+B_{2u}+B_{3u}$ \cite{I. S.
Smirnova}. In an infrared measurements on polycrystalline samples,
the absorption spectra are broadened and therefore it is difficult
to distinguish all the phonon modes as against measurements
carried out on single crystal samples \cite{A. Paolone-1}. The
infrared absorption peak evident at $\sim400$ and $600$ $cm^{-1}$
may be ascribed to the $B_{u}$ symmetry bending and stretching
mode, respectively \cite{A. Congeduti}. According to Kim et al.,
bending mode is the motion of Mn and O ions located along a
particular direction against the other oxygen ions in a plane
perpendicular to the direction. While, the peak attributable to
the stretching of Mn-O bonds corresponds to the internal motion of
Mn ions against the oxygen octahedron and is sensitive to Mn-O
bond lengths. At 300K, the infrared absorption peak at
$\sim600cm^{-1}$ is screened by a large background. Below 300K,
the background is reduced and the absorption peak shows a strong
temperature dependence, with built-up in intensity of stretching
mode peak. Previously reported neutron diffraction results on
these compounds show a large change in the apical and equatorial
bond lengths, following the charge and orbitally ordered
transition $(T_{CO}\approx250K)$ \cite{I. Dhiman}. The difference
between Mn-O bond lengths ($Mn-O_{1}$ along b axis (apical),
$Mn-O_{21}$ and $Mn-O_{22}$ in ac plane (equatorial)) increases as
temperature is reduced, which correlate with the stretching mode
vibrations of $Mn-O$ bonds as described by Smirnova \cite{I. S.
Smirnova}. On lowering of temperature to 200K, an additional mode
at $\sim510cm^{-1}$ appears in the form of shoulder in main
absorption peak and this becomes pronounced on reduction of
temperature. This mode has been observed in other charge ordered
manganites and has been identified with the structural
transformation of orthorhombic structure in \textit{Pnma} space
group to charge and orbitally ordered monoclinic structure in
$P2_{1}/m$ space group \cite{P. Calvani,V. TaPhuoc,I. S.
Smirnova}. This behavior is in agreement with our previously
reported neutron diffraction measurements on these samples,
wherein the structural transformation from orthorhombic to charge
and orbitally ordered monoclinic phase is observed while cooling.
Contrastively, in compounds away from half doping, such as
$La{}_{0.33}Ca_{0.67}MnO_{3}$ and $La{}_{0.67}Ca_{0.33}MnO_{3}$
systems, the \textit{Pnma} space group is preserved down to lowest
temperature, and shoulder like feature is absent in these samples,
indicating that the mode at $510cm^{-1}$ arises due to the
lowering of symmetry as a result of charge ordering \cite{P.
Calvani,M. Premila}. Further, on reducing temperature the
absorption background is seen to deepen. According to Calvani et
al. such gradual deepening of background with reducing temperature
for $La{}_{0.5}Ca{}_{0.5}MnO_{3}$ compound provides an evidence
for increasing localization of charge carriers \cite{P. Calvani}.
This is in agreement with the observation of sharp increase in
resistivity below $T_{CO}$ in Sr doped samples \cite{I. Dhiman}.
In contrast, the absorption background is seen to increase and the
associated stretching mode at $590cm^{-1}$ is diminished in the
ferromagnetically ordered $La{}_{0.67}Ca{}_{0.33}MnO_{3}$ system
exhibiting an insulator to metal transition \cite{M. Premila}. The
infrared active mode at 190$cm^{-1}$ ascribed to external mode was
not observed in the present infrared study due to the limited
range of instrument between 400 to 4000$cm{}^{-1}$ \cite{K. H.
Kim}. The external mode at 190$cm^{-1}$ represents a vibrating
motion of the La (Ca) ions against the $MnO{}_{6}$ octahedra. In
$La{}_{0.5}Ca{}_{0.2}Sr{}_{0.3}MnO_{3}$ (x = 0.3) sample with
coexisting CE - and A - type antiferromagnetic ordering, the
optical density ($O_{d}(\omega)$) shows similar temperature
dependence between 100 and 300K, as observed in x = 0.1 sample. In
x = 0.3 compound at 100K, the infrared absorption peak
attributable to the stretching mode vibrations of $MnO{}_{6}$
octahedra is observed at $620cm^{-1}$ and is accompanied by a
shoulder at $528cm^{-1}$, in association with the charge ordering
transition. The increase in $<r_{A}>$ as a result of Sr doping
leads to hardening of stretching mode.

\begin{figure}
\noindent \includegraphics{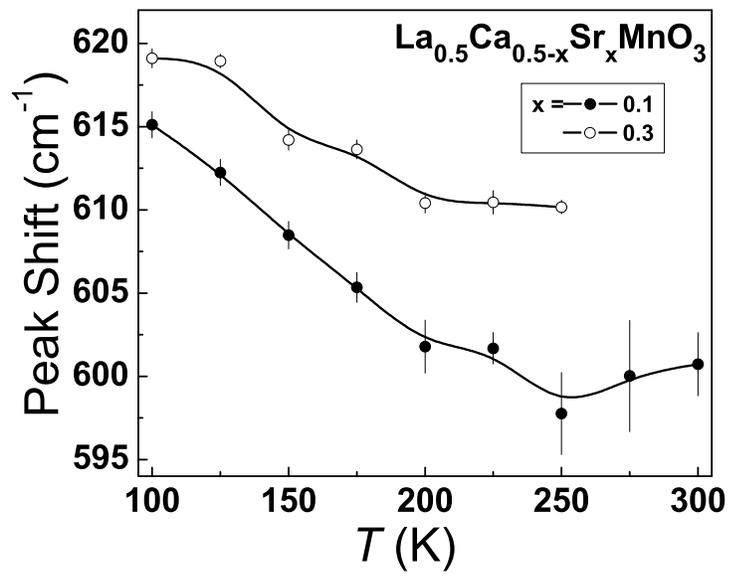}

\noindent \caption{\label{Frequency-shift_Sr01}Frequency shift of the stretching mode
peak at $\sim600-620cm^{-1}$ as a function of temperature for $La{}_{0.5}Ca{}_{0.5-x}Sr{}_{x}MnO_{3}$
(x = 0.1 and 0.3) samples.}

\end{figure}

\noindent Figure \ref{Frequency-shift_Sr01} displays frequency of
stretching mode as a function of temperature in x = 0.1 and 0.3
samples. The frequencies are obtained by fitting the absorption
curves to Lorentzian functions with the central frequency, half
width and area under the peak as fitting parameters. At
temperatures above the charge ordering transition a single
Lorentzian function accounting for the stretching mode is used,
while at low temperatures two Lorentzian functions assigned to
stretching mode and the additional mode are used. It is observed
that in both these samples a rapid shift in peak position is
observed below 200K close to $T_{N}$. Around $T_{CO}\approx250K$,
no significant change is observed. This indicates coupling of the
magnetic ordering with phonon modes. The temperature dependence of
frequency of stretching mode in x = 0.3 sample displays behavior
similar to x = 0.1 sample, as shown in figure
\ref{Frequency-shift_Sr01}. In x = 0.3 compound, above 250K, the
estimation of frequency shift is difficult due to the significant
broadening of the peak. The stretching mode vibration increases at
the antiferromagnetic transition temperature $T_{N}$ = 150K, as
observed in x = 0.1 composition. Similar behavior of stretching
mode frequency around the ferromagnetic transition and has been
reported for $La{}_{0.67-x}Pr_{x}Ca_{0.33}MnO_{3}$ compounds
\cite{Z. M. Lu}. The stretching mode frequency at 100K increases
from $\approx$615(1) $cm^{-1}$ to 620(1)$cm^{-1}$ with increase in
x from 0.1 to 0.3, indicating the hardening of phonon mode. This
is similar to the earlier reported observation of reduction in
stretching mode frequency with increase in Ca substitution (with
reducing $<r_{A}>$) in $La_{1-x}Ca_{x}MnO_{3}$ \cite{L. Kebin}.
Theoretically, Lee and Min have examined the origin of phonon
hardening (shifting of absorption peak to higher frequency below
the ordering temperature) both in the charge ordered phase and the
metallic phase of manganites. For charge ordered systems the study
reveals that ordering of localized polarons is responsible for
such a behavior \cite{J. D. Lee,J. D. Lee-1}.

\begin{figure}
\includegraphics{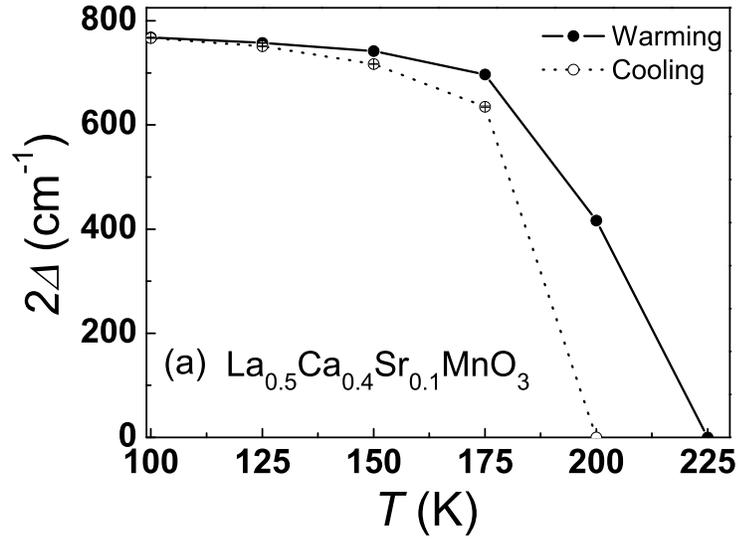}

\includegraphics{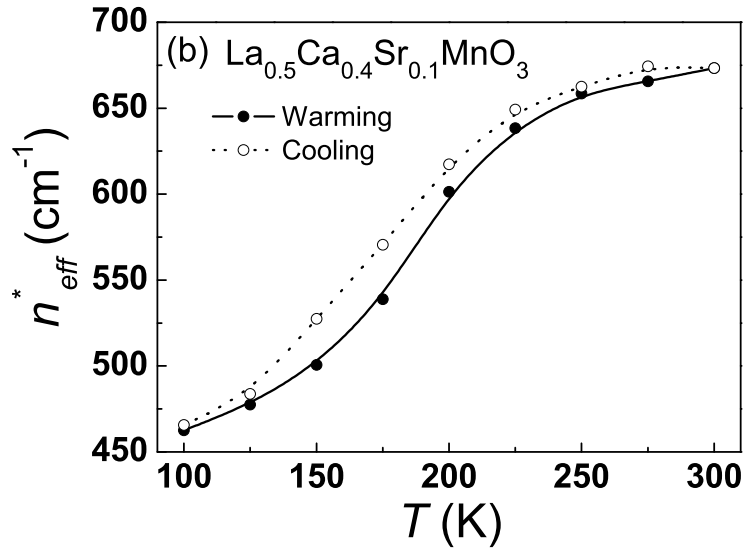}\caption{\label{Band gap_carriers} (a) the optical band gap $(2\Delta)$ and
(b) effective number of carriers $(n_{eff}^{*})$ as a function of
temperature for $La{}_{0.5}Ca{}_{0.4}Sr{}_{0.1}MnO_{3}$ sample for
warming (solid lines and full circles) and cooling (dotted lines and
open circles) cycles.}

\end{figure}

The insulating gap has been obtained by fitting the optical
density $O_{d}(\omega)$, shown in figure \ref{Optical-density
Sr01}, to the expression \cite{P. Calvani},

\begin{equation}
O_{d}(\omega)\varpropto[\omega-2\triangle(T)]^{1/2}\label{eq:optical
conductivity}\end{equation} for $\omega\geq2\triangle(T)$, where
$2\triangle(T)$ is the insulating gap. The fit to equation
\ref{eq:optical conductivity} for x = 0.1 sample is shown in
figure \ref{Optical-density Sr01} by dotted curves. The resulting
values of $2\triangle(T)$ as a function of temperature for both
heating and cooling cycles in x = 0.1 sample is shown in figure
\ref{Band gap_carriers}(a). As temperature is reduced the
insulating gap increases, and reaches a stable value for
$T\leq150K$. The temperature at which insulating gap shows a sharp
rise is identified with charge ordering transition temperature.
This is in agreement with that reported from $\rho(T)$ studies. In
addition, thermal hysteresis behavior in $2\triangle(T)$ for
heating and cooling cycles is clearly discernible near the
magnetic transition. This behavior arises due to the structural
transition accompanying the antiferromagnetic transition, a
characteristic feature occurring in $La{}_{0.5}Ca{}_{0.5}MnO_{3}$
system exhibiting CE-type antiferromagnetic state \cite{P.
Calvani,P. G. Radaelli}. For x = 0.1 sample, at 100K, the
insulating gap is $\sim727(1)cm^{-1}$. This value is in agreement
with the previously reported infrared absorption study on
$La{}_{0.5}Ca{}_{0.5}MnO_{3}$ compound yielding an insulating gap,
$2\triangle(T)\approx710cm^{-1}$ \cite{P. Calvani}. The variation
of effective carrier numbers as a function of temperature,
$n{}_{eff}^{*}(T)$, for x = 0.1 composition is shown in figure
\ref{Band gap_carriers}(b). This quantity is obtained by
integration of optical density $O_{d}(\omega)$ over a suitable
frequency range. The integration limits are the frequencies where
$O_{d}(\omega)$ does not change appreciably with temperature. We
have chosen $\omega{}_{1}=400cm^{-1}$ and
$\omega{}_{2}=1500cm^{-1}$. In terms of optical density the
expression for number of effective carriers $n{}_{eff}^{*}(T)$ is
given as \cite{P. Calvani},

\begin{equation}
n{}_{eff}^{*}(T)=\intop_{\omega_{1}}^{\omega_{2}}O_{d}(\omega)d\omega\varpropto n{}_{eff}(T)\label{eq:carrier density}\end{equation}

\begin{figure}
\noindent \includegraphics{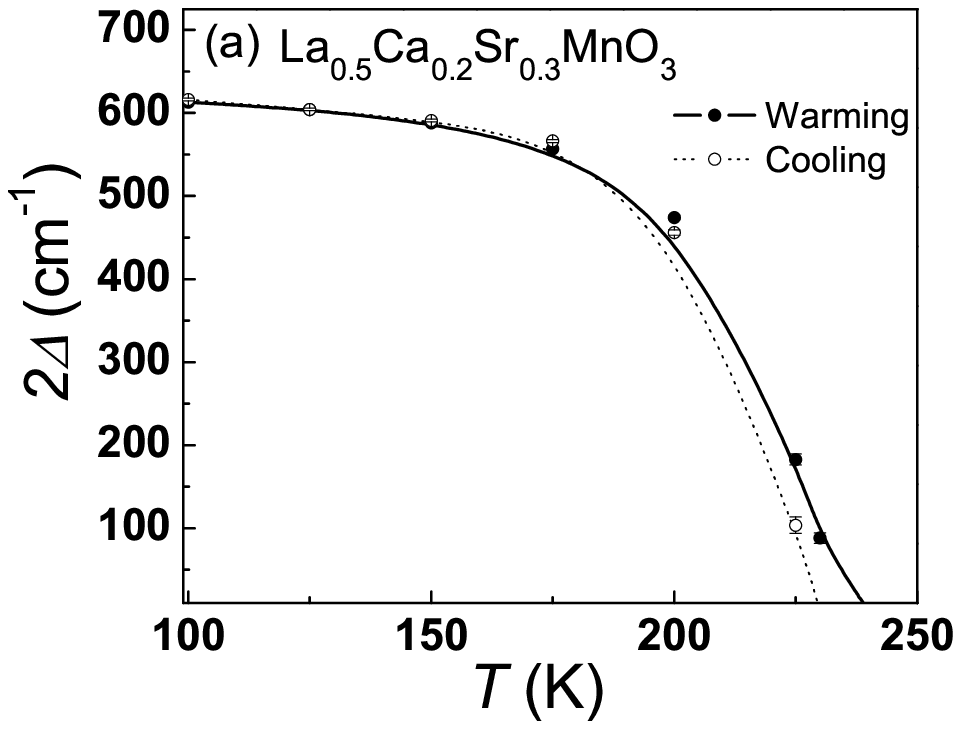}

\includegraphics{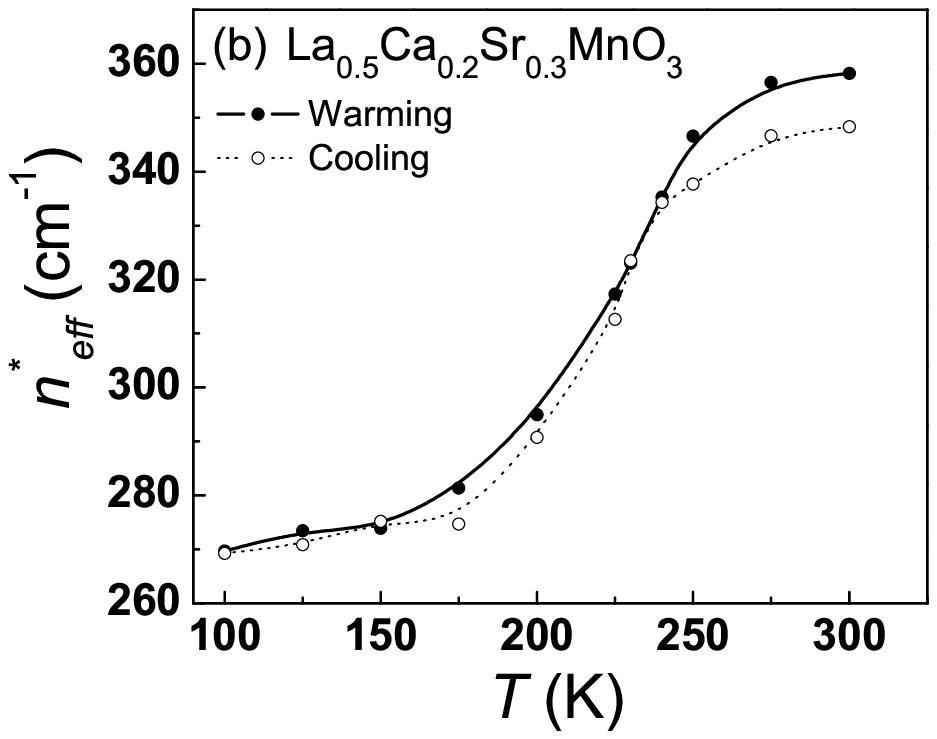}

\caption{\label{Band gap_carriers-Sr03} (a) the optical band gap $(2\Delta)$
and (b) effective number of carriers $(n_{eff}^{*})$ as a function
of temperature for $La{}_{0.5}Ca{}_{0.2}Sr{}_{0.3}MnO_{3}$ sample
for warming (solid lines and full circles) and cooling (dotted lines
and open circles) cycles.}

\end{figure}

\noindent As evident in figure \ref{Band gap_carriers}(b) at low
temperatures for large values of $2\triangle(T)$ (figure \ref{Band
gap_carriers}(a)), effective number of carriers are considerably
reduced. Above 200K $(\sim T_{CO})$, the band gap starts reducing,
which is consistent with an increasing effective number of charge
carriers. Besides, $n{}_{eff}^{*}(T)$ follows a thermal hysteresis
behavior, similar to that evident in $2\triangle(T)$. Similar to x
= 0.1 sample, from $O_{d}(\omega)$ data for x = 0.3 compound,
$2\triangle(T)$ and $n{}_{eff}^{*}(T)$ are obtained. Figure
\ref{Band gap_carriers-Sr03} (a) and (b) shows temperature
dependence of $2\triangle(T)$ and $n{}_{eff}^{*}(T)$ in x = 0.3
sample, respectively. Increase in $2\triangle(T)$ and correlated
reduction in $n{}_{eff}^{*}(T)$ in x = 0.3 Sr doped sample on
reducing temperature, similar to x = 0.1 composition, is in
agreement with the previously reported transport studies \cite{I.
Dhiman}. Although, in comparison to x = 0.1 compound, the width of
thermal hysteresis in $2\triangle(T)$ and $n{}_{eff}^{*}(T)$ is
considerably reduced. Also, with increasing Sr doping, the value
of $2\triangle(T)$ is reduced, which is in concurrence with
weakening of charge ordered insulating state at higher Sr doping.
Additionally, on lowering of temperature the absorption background
is observed to decrease. Similar correlation between resistivity
and $n{}_{eff}^{*}(T)$ has been observed in magnetically distinct
compounds such as $La{}_{0.7}Ca{}_{0.3}MnO_{3}$ and
$La{}_{0.75}Ca{}_{0.25}MnO_{3}$ \cite{A. Congeduti,K. H. Kim-1}.
These compounds upon cooling near ferromagnetic transition
temperature, show a transition from insulating to metallic state.
Infrared study on these systems reveals a reduction in number of
carriers with increasing temperature. There exists considerable
disparity in the insulating band gap $2\triangle(T)$ for charge
ordered manganites. Previous report on $La{}_{1-x}Ca{}_{x}MnO_{3}$
show that $2\triangle(T)$ values increase with x. At x = 0.5
compound, the gap is $\sim0.45eV$ ($3600cm^{-1}$) \cite{K. H.
Kim-2}. Another study on $La{}_{0.5}Ca{}_{0.5}MnO_{3}$ reports the
gap to be $320cm^{-1}$ \cite{A. P. Litvinchuk}. However, in the
$La_{5/8-y}Pr_{y}Ca_{3/8}MnO_{3}$ system, where the charge
ordering is known to be of $La{}_{0.5}Ca{}_{0.5}MnO_{3}$ type, the
insulating gap is $\sim0.38eV$ ($3065cm^{-1}$) \cite{H. J. Lee}.

The strength of electron - phonon coupling is related with the
ratio given as, $2\triangle/k_{B}T_{CO}$ \cite{C. A. Perroni}. In
x = 0.1 sample, the calculated value of $2\triangle/k_{B}T_{CO}$
is $\approx4.8$, which is reduced to 3.7 in x = 0.3 compound. This
indicates that as Sr doping increases the strength of electron -
phonon interaction is reduced. This suggests that enhanced
electron - phonon coupling interaction may be important for
stabilizing the charge ordered state. The obtained value of
$2\triangle/k_{B}T_{CO}$ is in agreement with
$La_{0.5}Ca_{0.5}MnO_{3}$ sample having value $\approx4.8$
\cite{P. Calvani}. Similar trend has been observed in
$La_{1-x}Ca_{x}MnO_{3}$ as a function of Ca doping, wherein
$2\triangle/k_{B}T_{CO}$ ratio is maximized for x = 0.50;
therefore favoring the enhancement of charge ordered state
\cite{K. H. Kim-2}. However, according to this study the deduced
ratio is of the order of $\approx$30 in x = 0.5 compound, which is
much higher than the values reported here for Sr doped samples and
previously reported study on $La_{0.5}Ca_{0.5}MnO_{3}$ \cite{P.
Calvani}.

\begin{figure}
\includegraphics{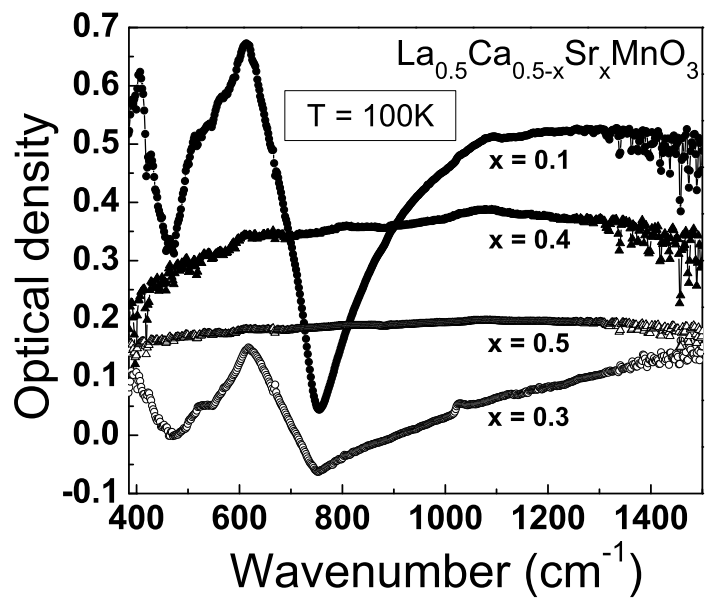}

\caption{\label{IR-spectra_x} Infrared absorption spectra for $La{}_{0.5}Ca{}_{0.5-x}Sr{}_{x}MnO_{3}$
(0.1 \ensuremath{\le} x \ensuremath{\le} 0.5) series at 100K.}

\end{figure}

At higher Sr doping in $La{}_{0.5}Ca{}_{0.1}Sr{}_{0.4}MnO_{3}$ (x
= 0.4), neutron diffraction study carried out on this compound
reveals that the antiferromagnetic spin structure changes from
CE-type to A-type and consequently the nature of orbital ordering
is modified\cite{I. Dhiman,I. Dhiman-1}. Further, transport
measurements have shown that at low temperatures the resistivity
of this compound is lowered as compared to the samples with
$x\leq0.3$ \cite{I. Dhiman}. The infrared absorption spectra in x
= 0.4 sample with A - type antiferromagnetic order, as shown in
figure \ref{IR-spectra_x}, is significantly different in
comparison to samples with $x\leq0.3$ . While the typical Mn-O
bond stretching mode with considerable weakening is still visible
around $\sim610cm^{-1}$, the shoulder like feature observed
in$x\leq0.3$ samples is no longer visible in x = 0.4 sample. As
discussed above, this feature has been ascribed to a transition to
charge ordered state. Therefore, disappearance of this feature may
be caused by the suppression of charge ordering in x = 0.4
compound. Similar behavior has been observed in compound
exhibiting CE - and A - type antiferromagnetic ordering. The
stretching and bending mode in the $Nd{}_{0.5}Sr{}_{0.5}MnO_{3}$
compound with CE - type antiferromagnetic ground state, exhibit
splitting below $T_{N}$, which is in contrast with
$Pr{}_{0.5}Sr{}_{0.5}MnO_{3}$ compound displaying A - type
antiferromagnetic state. Nevertheless, the background attributed
to localization of charge carriers still exhibits reduction while
cooling from 300K. This implies decrease in the number of
effective carriers with reducing temperature. However, the
estimated value of $n{}_{eff}^{*}(T)$ using equation
\ref{eq:carrier density} exhibits a nearly linear temperature
dependence, in contrast to compounds with $x\leq0.3$ . A fit to
equation \ref{eq:optical conductivity}, as performed in $x\leq0.3$
compounds, could not be carried out in x = 0.4 composition due to
significant weakening of the stretching mode peak. Interestingly,
in a previously reported optical conductivity measurements on
single crystal of $Pr{}_{0.5}Sr{}_{0.5}MnO_{3}$ compound, having
an A-type antiferromagnetic spin structure, the temperature
dependence of spectral weight (measured by the variation of
$n{}_{eff}^{*}(T)$) shows a strong correlation with the magnetic
phase transition \cite{J. H. Jung}. The infrared absorption
measurements as a function of temperature were also performed in x
= 0.5 composition, which display tetragonal crystal structure with
\textit{I4/mcm} space group and the long range ordered
ferromagnetic state at all temperatures below 310K. A broad hump
attributable to the presence of stretching mode at
$\sim600-620cm^{-1}$ is observed, exhibiting no significant change
as a function of temperature. Similar to samples with $x\leq0.3$ ,
the $n{}_{eff}^{*}(T)$ exhibits an increase with increasing
temperature. This provides a signature of reduction in optical
band gap with Sr doping, favoring delocalization of carriers.
Therefore, in the present study we observe that position and
intensity of the absorption peak is significantly affected as a
function in Sr doping in $La_{0.5}Ca_{0.5-x}Sr_{x}MnO_{3}$ system.
Similar behavior has been observed in polycrystalline
$La_{1-x-y}R_{y}Ca_{x}MnO_{3-\delta}$ manganites, wherein the
intensity and position of the absorption peak are influenced by
the modification in local lattice distortions as a function of
doping \cite{L. Kebin}.

Thus, with increasing Sr doping (increasing A-site ionic radii
$<r_{A}>$ and disorder $\sigma^{2}$) the stretching mode peak is
weakened and delocalization tendencies are favored due to increase
in $<Mn-O-Mn>$ bond angles towards $180^{\circ}$ and therefore
enhancing the one electron bandwidth. These structural
modification leads to collapse of insulating gap and favor hopping
and therefore result in charge delocalization. Effect similar to
internal pressure (caused by increasing $<r_{A}>$) has been
achieved on application of external pressure \cite{I. Dhiman-2}.
Previous pressure dependent infrared absorption studies at room
temperature on $La_{0.75}Ca_{0.25}MnO_{3}$ system reveals
reduction in optical band gap as pressure is increased \cite{A.
Sacchetti,A. Congeduti,A. Sacchetti-1}.

\section{Conclusion}

We have investigated the half doped $La_{0.5}Ca_{0.5-x}Sr_{x}MnO_{3}$
$(0.1\leq x\leq0.5)$ manganites with different magnetic ground states
by infrared absorption measurements. In samples with CE-type antiferromagnetic
ground states with $x\leq0.3$, the stretching mode vibration of $MnO_{6}$
octahedra at $\sim615-620cm^{-1}$ in addition to a mode at $\sim510-525cm^{-1}$
attributable to charge ordered phase is evident for these samples.
The obtained values of insulating band gap $2\triangle(T)$ and effective
number of charge carriers $n{}_{eff}^{*}(T)$ in x = 0.1 and 0.3 samples,
exhibit decrease on lowering of temperature. At 100K, the values of
$2\triangle(T)$ in x = 0.1 and 0.3 samples are $\sim727(1)$ and
$605(1)cm^{-1}$, respectively. For Sr doping with $x\geq0.4$, considerable
weakening of stretching mode is seen, while shoulder ascribed to charge
ordered state is absent.

\newpage{}

\noindent \listoffigures
\pagebreak{}

\noindent %

\noindent %

\noindent %

\noindent %

\noindent %

\end{document}